\documentstyle[12pt,epsf]{article}
\begin{document}
\hoffset=-1.2cm
\hsize=16cm
\vsize=24cm

\begin{flushright}
DTP-94/48\\
June, 1994\\
\end{flushright}

{\centerline{\bf {GAUGE INDEPENDENT CHIRAL SYMMETRY BREAKING}}}
\vskip 2mm
{\centerline{\bf {IN QUENCHED QED}}}
\vskip 2cm
\baselineskip=7mm
{\centerline{{\bf{A. Bashir and M.R. Pennington}}}}
\vskip 5mm
{\centerline{Centre for Particle Theory,}}
{\centerline{University of Durham}}
{\centerline{Durham DH1 3LE, U.K.}}
\vskip 2cm
{\centerline {ABSTRACT}}
{\noindent In quenched QED we construct a non-perturbative
fermion-boson vertex that ensures the
fermion propagator satisfies the Ward-Takahashi identity, is multiplicatively
renormalizable, agrees with perturbation theory for weak couplings and has a
critical coupling for dynamical mass generation that is strictly
gauge independent.
This is in marked contrast to the {\it rainbow} approximation in which the
critical coupling changes by 50\% just between the Landau and Feynman gauges.
The use of such a vertex should lead to a more believable study of mass
generation.}
\vfil\eject
\vskip 2cm
\section{Introduction}
\baselineskip=9.5mm

The standard model is highly successful in collating experimental
information on the basic forces. Yet, its key parameters, the masses
of the quarks and leptons, are theoretically undetermined. In the
simplest version of the model, these masses are specified by the
couplings of the Higgs boson, couplings that are in turn undetermined.
However, it could be that it is the dynamics of the fundamental gauge
theories themselves that generate the masses of all the matter fields.
To explore this possibility, the favourite starting point is to
consider quenched QED [1-9]
as the simplest example of a
gauge theory and
study the behaviour of the fermion propagator. Then in the {\em rainbow}
approximation, it is well-known that the fermion field can have a
dynamically generated mass if the interaction is strong enough, i.e.
the coupling, $\alpha$, is larger than some critical value,
$\alpha_{c}$. This critical coupling marks a change of phase and so
its value should be gauge independent. Unfortunately, the {\em rainbow}
approximation allows a far from gauge invariant treatment \cite{Aoki,Kamli}.
The
purpose of this paper is to construct a non-perturbative fermion-boson
interaction that respects the Ward-Takahashi identity, ensures the fermion
propagator is
multiplicatively renormalizable, agrees with perturbation theory
when $\alpha \ll 1$, and possesses a gauge independent critical
coupling.

 The {\em gauge technique} of Salam,  Delbourgo and collaborators
\cite{Salam} was developed to
 solve essentially such constraints.  However, despite  formal results
 on the first two of these~\cite{Delbourgo,GT}, their expression in terms of
the spectral
 representation for the fermion propagator has proved
 difficult in practical calculations of the
 fermion propagator, for example,~\cite{Rem1}.
  Consequently, we develop an explicit
 construction procedure amenable to straightforward computation.

We start with the Schwinger-Dyson equation  ~  for the fermion propagator. This
non-linear
integral equation encodes all we can know about the fermion
propagator. To be able to consider this equation alone of the infinite
set of Schwinger-Dyson equations -- one for each Green's function --
we must make an ansatz for
the full fermion-boson vertex. Quite generally, this vertex can be
regarded as the sum of two components~: the longitudinal and transverse
parts.  The well-known Ward-Takahashi identity constrains the longitudinal
part. How to
fulfill this constraint in a manner free of kinematic singularities
has been solved some time ago by Ball and Chiu \cite{BC}. That multiplicative
renormalizability  constrains
the transverse vertex has also been known for some time \cite{MR,GT}. However,
it is more recently that Curtis and one of the present authors \cite{CP}
explicitly constructed a simple form (perhaps the simplest possible
form) to ensure the multiplicative renormalizability  of the fermion
propagator. This ansatz is
called the {\em CP} vertex.

Subsequent study has shown that with this vertex the fermion propagator
still has the possibility of a chiral symmetry breaking phase \cite{ABGPR}.
Moreover, in dramatic contrast to the {\em rainbow} approximation, the
critical coupling required is only very weakly gauge dependent in the
neighbourhood of the Landau gauge. However weak this variation, any
gauge dependence shows that the {\em CP} vertex cannot be the exact choice.
Here, we determine the constraints on the full fermion-boson vertex
that ensures gauge covariance for the fermion propagator and exact
gauge independence for the critical coupling. The resulting vertex
involves two unknown functions $W_1$ and  $W_2$, which each satisfy a
sum rule and a constraint on their derivatives. Any choice of these
fulfills our fundamental constraints as long as it correctly matches
onto perturbation theory. This construction builds on the $CP$ vertex,
extending
the work of Dong et al. \cite{Dong}.  Though the discussion in Sect. 2
of how to ensure the gauge
covariance of the wavefunction renormalization of the fermion propagator
is very close to that of Dong et al. \cite{Dong}, to make the extension
to the gauge independence of the critical coupling clear, we have
given all the details of our formulation  making our construction in Sect. 3.
self-contained.

In general, only the position of the pole in a propagator has to be
gauge independent. At that value of the momentum, when $p^2=m^2$ in
Minkowski space, (or equivalently at $p^2=-m^2$ in the Euclidean space
in which we work) the fermion mass function has to be independent of
the gauge. Atkinson and Fry \cite{Fry} proved this independence
follows from the
Ward-Takahashi identities. However, at the critical coupling for
dynamical mass generation, multiplicative renormalizability imposes
such a simple form on the mass function that this whole function
becomes gauge independent. This is embodied in our construction.

Our results have to be compared with earlier work.
  For example, Rembiesa \cite{Rembiesa} and Haeri \cite{Haeri}, using
the previously mentioned {\em gauge technique},
 construct fermion-boson vertices that
make the fermion propagator itself gauge independent.
 This is, of course, at variance with its behaviour in
perturbation theory and consequently with
the renormalization group in the weak coupling limit.
Rembiesa \cite{Rembiesa} then went on to find that the
 critical coupling for mass generation
with such a vertex is strongly gauge dependent, being given by
 $\alpha_c = \pi/(3+\xi)$.  In complete contrast, Kondo \cite{Kondo}
finds a gauge independent coupling as here,
 but at the expense of using a vertex
that has singularities.  The construction presented here aims to overcome these
deficiencies.


\section{The Fermion Equation}
\baselineskip=9.5mm

The Schwinger-Dyson equation for the fermion propagator, $S_{F}(p)$,
in QED with a bare
coupling, $e$, is displayed in Fig 1, and is given by:
\begin{eqnarray}
     iS_{F}^{-1}(p)\,=\,iS_{F}^{0^{-1}}(p)\,-\,e^2\int \frac{d^4k}{(2\pi)^4}\,
    \gamma^{\mu}\,S_{F}(k)\, \Gamma^{\nu}(k,p)\,\Delta_{\mu\nu}
    (q) \quad ,
\end{eqnarray}
where  $q=k-p$ and $S_{F}(p)$ can be expressed in terms of two Lorentz scalar
functions, $F(p^2)$ the wavefunction renormalization and
${\cal M}(p^2)$ the mass function, so that
\begin{eqnarray*}
    S_{F}(p)&=& \frac{F(p^2)}{\not \! p- {\cal M}(p^2)}\qquad .
\end{eqnarray*}
The bare propagator $ S_{F}^{0}(k)= 1/ (\rlap / p - m_{0}) $, where $m_0$
is the constant (bare) mass. In quenched QED, the photon propagator is
unrenormalized and so is given by its bare form~:
\begin{eqnarray*}
 \Delta_{\mu\nu}(q) \equiv \Delta_{\mu\nu}^{0}(q) =\,\frac{1}{q^2}\ \left(
    g_{\mu\nu}+(\xi-1)\frac{q_{\mu}q_{\nu}}{q^2}\ \ \right) \equiv
     \Delta_{\mu\nu}^{T}(q)+\xi\frac{q_{\mu}q_{\nu}}{q^4} \qquad ,
\end{eqnarray*}
where the transverse part, $\Delta_{\mu\nu}^{T}(q)$, is defined by
this equation and $\xi$ is the standard covariant
gauge parameter. $\Gamma^{\mu}(k,p)$ is the full fermion-boson vertex
that must satisfy the Ward-Takahashi identity
\begin{eqnarray}
    q^{\mu}\Gamma_{\mu}(k,p)= S_{F}^{-1}(k)-S_{F}^{-1}(p) \qquad.
\end{eqnarray}

We can simplify Eq. (1) by making use of the Ward-Takahashi identity,
 Eq. (2)~:
\begin{eqnarray}
\nonumber
     S_{F}^{-1}(p)\,=\,S_{F}^{0^{-1}}(p)&+&ie^2\int \frac{d^4k}{(2\pi)^4}\
     \frac{1}{q^2}\
     \gamma^{\mu}\,S_{F}(k)\, \Gamma^{\nu}(k,p)\
        \Delta_{\mu\nu}^{T}(q)    \\ \nonumber
   &+& ie^2 \xi \int \frac{d^4k}{(2\pi)^4}\ \frac{\not \! q}{q^4}\
\\
   &-& ie^2 \xi \int \frac{d^4k}{(2\pi)^4}\
     \frac{\not \! q}{q^4}\  S_{F}(k) S_{F}^{-1}(p)\quad .
\end{eqnarray}
The third term on the right vanishes \footnote {This was
not noted in Ref. \cite{ABGPR} as pointed out in \cite{Dong},
 who better remembered
Ref. 20 of \cite{BurdenCP} than the authors !}, as it is an odd integral, and
we
are left with
\vspace {2mm}
\begin{eqnarray}
\nonumber
     S_{F}^{-1}(p)\,=\,S_{F}^{0^{-1}}(p)&+&ie^2\int \frac{d^4k}{(2\pi)^4}\
     \frac{1}{q^2}\
     \gamma^{\mu}\,S_{F}(k)\, \Gamma^{\nu}(k,p)\
       \Delta_{\mu\nu}^{T}(q)  \\
   &-& ie^2 \xi  \int \frac{d^4k}{(2\pi)^4}\
    \frac{\not \! q}{q^4}\  S_{F}(k) S_{F}^{-1}(p)\qquad .
\end{eqnarray}

To solve this equation we must make an ansatz
for the full vertex, $\Gamma^{\mu}(k,p)$.
 Our aim is to construct a vertex that automatically embodies as much
of the physics of the interaction as possible.
Following Ball and Chiu \cite{BC}, we first
write the vertex as a sum of longitudinal
and transverse components~:
 \begin{eqnarray}
      \Gamma^{\mu}(k,p)=\Gamma^{\mu}_{L}(k,p)+\Gamma^{\mu}_{T}(k,p)\qquad .
\end{eqnarray}
 To satisfy Eq. (2) in a manner free of kinematic singularities,
which in turn ensures the Ward identity is fulfilled, we have
(following Ball and Chiu)~:
\begin{eqnarray}
\Gamma^{\mu}_{L}(k,p)= a(k^2,p^2) \gamma^{\mu}
                       + \mbox{} b(k^2,p^2) (\not \! k + \not \! p)
                           (k+p)^{\mu}
                       - \mbox{} c(k^2,p^2) (k+p)^{\mu}
\end{eqnarray}
 where,
\begin{eqnarray}\nonumber
  a(k^2,p^2)&=&\frac{1}{2}\ \left( \frac{1}{F(k^2)}\ +
  \frac{1}{F(p^2)}\  \right) \qquad \qquad , \\
  b(k^2,p^2)&=&\frac{1}{2}\ \left(
  \frac{1}{F(k^2)} - \frac{1}{F(p^2)}\ \right) \frac{1}{k^2 - p^2}\quad ,  \\
\nonumber
  c(k^2,p^2)&=&  \left( \frac{{\cal M}(k^2)}{F(k^2)}\
  - \frac{{\cal M}(p^2)}{F(p^2)}\ \right) \frac{1}{k^2 - p^2} \quad ,
\end{eqnarray}
and
\begin{eqnarray}
        q_{\mu} \Gamma^{\mu}_{T}(k,p)=0 \hspace{5 mm}\ , \hspace{5
        mm}\ \Gamma^{\mu}_{T}(p,p)=0 \quad .
\end{eqnarray}
Ball and Chiu wrote down a set of 8 basis vectors $T_{i}^{\mu}(k,p)$
for the transverse part \cite{BC}, that ensures these conditions, Eq.
(8), are fulfilled~:
\vspace {2mm}
\begin{eqnarray}
   \Gamma^{\mu}_{T}(k,p)= \sum_{i}^8 \tau_{i}(k^2,p^2,q^2)T_{i}^
     {\mu}(k,p)
\end{eqnarray}
\vspace {2mm}
provided that in the limit $k\rightarrow p $, $\tau_{i}(p^2,p^2,0)$
are finite. Our aim is to determine the full vertex by requiring the
multiplicative renormalizability of the fermion propagator and the
gauge independence of the chiral symmetry breaking
phase transition. Since the longitudinal part of this vertex is
specified, Eq. (6), this amounts to determining the transverse
part and hence the $\tau_{i}$ of Eq. (9).
Of the eight basis vectors, $T_i^{\mu}$,
four have even numbers of gamma matrices and four have odd numbers.

It is here that we make three simplifying assumptions.
 Firstly, we demand that a chirally-symmetric solution
 should be possible when the bare mass is zero, just as in perturbation theory.
 This is most easily accomplished if
  only those transverse vectors with odd numbers of
  gamma matrices contribute to $\Gamma_T^{\mu}(k,p)$.
  Then the sum in Eq. (9) involves just $i=2,3,6$ and 8.
   The corresponding vectors are~:
\vspace {2mm}
\begin{eqnarray}\nonumber
   T_{2}^{\mu}(k,p)&=& (p^{\mu} k\cdot q - k^{\mu}  p\cdot q) ( \not \! k
            +\not \! p)  \hspace{14mm},
              \\ \nonumber
   T_{3}^{\mu}(k,p)&=& q^{2} \gamma^{\mu}-q^{\mu} \not \! q  \hspace{37mm} ,
              \\  \nonumber
   T_{6}^{\mu}(k,p)&=& \gamma^{\mu} (k^{2} - p^{2})
                         -( k+p)^{\mu} ( \not \! k -\not \! p )  \quad ,
                                \\
   T_{8}^{\mu}(k,p)&=& -\gamma^{\mu} p^{\nu} k^{\rho} \sigma_{\nu\rho}
                       + p^{\mu} \not \! k - k^{\mu} \not \! p \; \qquad,
\end{eqnarray}
\vspace {2mm}
 where
$\sigma_{\mu\nu}= {1\over 2}\ [\gamma_{\mu},\gamma_{\nu}] \, .$

  The second assumption is  that the functions, $\tau_{i}$,
  multiplying the transverse
vectors, Eq. (9), only depend on $k^2$
and $p^2$, but {\bf not} $q^2$. This allows the angular integrations in
Eqs. (1,4) to be performed. Thirdly, we assume that, in the Landau
gauge, the transverse component of the vertex vanishes. This is
motivated by its large momentum behaviour in perturbation theory.
There for $k^2\gg p^2$, the leading logarithmic behaviour is \cite{CP}~:
\vspace {2mm}
\begin{eqnarray}
\Gamma_T^{\mu}(k,p)\,\simeq\,-\, \frac {\alpha\xi}{8\pi} \
                             \,\ln{k^2\over{p^2}}\,
\left[ \gamma^{\mu}\,-\,{k^{\mu} \rlap / k \over{k^2}} \right]\quad ,
\end{eqnarray}
\vspace {2mm}
where as usual $\alpha = e^2/4\pi$.

The fermion propagator is determined by the two functions $F(p^2)$ and
${\cal M}(p^2)$. We can project out equations for these by taking the
trace of Eq. (4) having multiplied by $\not \! p$ and 1 in turn.

\vfil\eject

\noindent On
Wick rotating to Euclidean space,
\begin{eqnarray}
\nonumber
 \frac{1}{F(p^2)}\,=\; 1\;- &&\frac{\alpha}{4\pi^3} \frac{1}{p^2} \int
                          d^4k \,
                          \frac{F(k^2)}{k^2+{\cal M}^2(k^2)}\
                          \frac{1}{q^2}\        \\   \nonumber
                   & &  \Bigg\{ \hspace{4 mm}\vspace{5 mm} a(k^2,p^2)
                         \left[-2k\cdot  p - \frac{1}{q^2} \left(
                       -2k^2p^2 + (k^2+p^2)k\cdot  p \right) \right]
                          \\  \nonumber
                   & &+ \hspace{4 mm}b(k^2,p^2)
                        \left[2k^2 p^2+(k^2+p^2)k\cdot  p
                          - \frac{1}{q^2}\ (k^2-p^2)^2 k\cdot  p
                        \right]  \\  \nonumber
                   & &+ \hspace{4 mm} {\cal
                   M}(k^2)c(k^2,p^2)\left[p^2+k\cdot p
                       - \frac{1}{q^2}\ (k^2-p^2)(k.p-p^2)  \right]
                    \\  \nonumber
                   & &-\frac{\xi}{q^2 F(p^2)}\ \left[
                       p^2 (k^2-k\cdot p) + {\cal M}(k^2)
                       {\cal M}(p^2)(k\cdot p-p^2)
                       \right]      \\ \nonumber
                   & &+\hspace{4mm}\tau_{2}(k^2,p^2)
                       \left[(k^2+p^2)(  k^2p^2
                      - (k\cdot p)^2 )  \right]   \\  \nonumber
                   & &+\hspace{4mm}\tau_{3}(k^2,p^2)
                       \left[-2k^2p^2 +3(k^2+p^2)k\cdot p
                      -4(k\cdot p)^2   \right]   \\  \nonumber
                   & &+\hspace{4 mm}\tau_{6}(k^2,p^2)
                       \left[(k^2-p^2)3k\cdot p  \right] \\
                   & &+\hspace{4 mm}\tau_{8}(k^2,p^2)
                       \left[-2k^2 p^2 +2(k\cdot p)^2 \right] \Bigg\}
\end{eqnarray}
\baselineskip=9.5mm
and
\begin{eqnarray}
\nonumber
 \frac{{\cal M}(p^2)}{F(p^2)}\ =\; m_0\; -&&\frac{\alpha}{4\pi^3} \int
                          d^4k \,
                          \frac{F(k^2)}{k^2+{\cal M}^2(k^2)}\
                          \frac{1}{q^2}\        \\   \nonumber
               &&\Bigg\{ -\,a(k^2,p^2) {\cal M}(k^2)
                         \left[ 3 \right] \\ \nonumber
                   &-& b(k^2,p^2) {\cal M}(k^2)
                        \left[ (k+p)^2 - \frac{1}{q^2}\ (k^2-p^2)^2
                         \right] \\
                          \nonumber
                   &+& c(k^2,p^2) \left[
                        (k^2+k\cdot p)- \frac{1}{q^2}\ (k^2-p^2) (k^2-
                        k\cdot p)  \right] \\  \nonumber
                   &-& \frac{\xi}{q^2 F(p^2)}\
                       \left[ {\cal M}(p^2) (k^2-k\cdot p) - {\cal M}(k^2)
                        (p\cdot k-p^2) \right] \\  \nonumber
                   &+&\tau_{2}(k^2,p^2) {\cal M}(k^2)
                       \left[ -2k^2p^2 + 2(k\cdot p)^2 \right]
                   \,+\,3\,\tau_{3}(k^2,p^2) {\cal M}(k^2)  \\
                   &+&\tau_{6}(k^2,p^2) {\cal M}(k^2)
                       \left[ 3(k^2-p^2)  \right] \Bigg\}\qquad .
\end{eqnarray}
 We are only interested in solving
 this equation when the bare mass, $m_0$ is zero.
 One solution of the mass equation, Eq. (13),
 is, as anticipated, ${\cal M}(p^2)=0$.
 We first consider the wavefunction renormalization,
  $F(p^2)$, in this case.
\vfil\eject

\noindent Carrying
out the angular integrations in Euclidean space gives~:
\vspace {2mm}
\begin{eqnarray}
 \nonumber \frac{1}{F(p^2)}\ = 1\, &+&\frac{\alpha\xi}{4\pi}\
                          \int_{p^2}^{\Lambda^2}
                          \frac{dk^2}{k^2}\
                          \frac{F(k^2)}{F(p^2)}\     \\  \nonumber
                   &-& \frac{3\alpha}{16\pi}\
                         \vspace{5 mm} \int_{0}^{p^2}\frac{dk^2}{p^2}\
                        \frac{k^2}{p^2}\  \frac{k^2+p^2}{k^2-p^2}\
                         \left( 1- \frac{F(k^2)}{F(p^2)}\ \right)  \\
                          \nonumber
                   &-& \frac{3\alpha}{16\pi}\
                         \vspace{5 mm} \int_{p^2}^{\Lambda^2}
                         \frac{dk^2}{k^2}\
                         \frac{k^2+p^2}{k^2-p^2}\
                         \left( 1- \frac{F(k^2)}{F(p^2)}\ \right)  \\
                         \nonumber
                   &-& \frac{\alpha}{8\pi}\
                         \vspace{5 mm} \int_{0}^{p^2}\frac{dk^2}{p^2}\
                        \frac{k^2}{p^2}\ F(k^2) K_{1}(k^2,p^2)  \\
                   &-& \frac{\alpha}{8\pi}\
                         \vspace{5 mm} \int_{p^2}^{\Lambda^2}\frac{dk^2}{k^2}\
                          F(k^2) K_{2}(k^2,p^2)\qquad ,
\end{eqnarray}
\vspace {2mm}
where
\vspace {2mm}
\begin{eqnarray}
\nonumber
       K_{1}(k^2,p^2)&=&(k^2-3p^2) \left[ \tau_{3}(k^2,p^2) +
        \tau_{8}(k^2,p^2) - \frac{1}{2}\ (k^2+p^2 )\, \tau_{2}(k^2,p^2)
       \right]         \\
      && \hspace{20mm} +\, 3(k^2-p^2) \,\tau_{6}(k^2,p^2)         \\ \nonumber
       K_{2}(k^2,p^2)&=&(p^2-3k^2) \left[ \tau_{3}(k^2,p^2) +
        \tau_{8}(k^2,p^2) - \frac{1}{2}\ (k^2+p^2 ) \,\tau_{2}(k^2,p^2)
       \right]         \\
      &&\hspace{20mm} +\, 3(k^2-p^2) \,\tau_{6}(k^2,p^2)\qquad .
\end{eqnarray}
\vspace {2mm}
The following treatment turns out to be
very close to that of Dong et al. \cite{Dong}, in
a more suitable form for our extension to dynamical mass generation.
It is convenient to define the combination $\overline{\tau}$ of $\tau_{2},
\tau_{3}$ and $\tau_{8}$,
\begin{eqnarray}
    \overline{\tau}(k^2,p^2)&=&\tau_{3}(k^2,p^2) + \tau_{8}(k^2,p^2) -
                  \frac{1}{2}\ (k^2+p^2)\, \tau_{2}(k^2,p^2)\quad .
\end{eqnarray}
Then
\begin{eqnarray}
       K_{1}(k^2,p^2)&=&(k^2-3p^2) \, \overline{\tau}(k^2,p^2)
      + 3(k^2-p^2)\, \tau_{6}(k^2,p^2)         \\
       K_{2}(k^2,p^2)&=&(p^2-3k^2) \, \overline{\tau}(k^2,p^2)
       + 3(k^2-p^2)\, \tau_{6}(k^2,p^2)\quad ,
\end{eqnarray}
which can be re-expressed in terms of
functions with definite symmetry properties when $k \leftrightarrow p$.
Thus
\begin{eqnarray}
   K_{1}(k^2,p^2)= h_s(k^2,p^2) +  h_a(k^2,p^2) \\
   K_{2}(k^2,p^2)= h_s(k^2,p^2) -  h_a(k^2,p^2)
\end{eqnarray}
 where $h_s(k^2,p^2)$ and $h_a(k^2,p^2)$ are symmetric and
antisymmetric respectively under the interchange of $k$ and $p$,
\begin{eqnarray}
 h_s(k^2,p^2)&=&-(k^2+p^2)\,\overline{\tau}(k^2,p^2) +
        3(k^2-p^2)\,\tau_{6}(k^2,p^2) \\
 h_a(k^2,p^2)&=&2(k^2-p^2)\,\overline{\tau}(k^2,p^2)
\end{eqnarray}
As discussed in \cite{Brown,CP,ABGPR}, multiplicative renormalizability
requires that the solution of this integral
equation for the wavefunction renormalization, $F(p^2)$, must be of
the form,
\begin{eqnarray}
    F(p^2)= A\left( p^2 \right)^\nu
\end{eqnarray}
As shown in \cite{Dong}, gauge covariance requires, $ \nu= \alpha \xi / 4
\pi\ $.
Burden and Roberts \cite{Burden} noted numerically that the fermion equation
with the simple
{\em CP}-vertex correctly generates this behaviour, even though the authors
of Ref. \cite{CP2,ABGPR} found  $ \nu= 2\alpha \xi / (8\pi  + \alpha \xi) $ as
a result of not imposing translational invariance on their loop
integrations, as discussed earlier.

This simple power behaviour is generated by the 1 and the first integral on
the right hand side of Eq. (12). This requires, as noted in Refs.
\cite{CP2,Dong},
a cancellation among the remaining integrals. Thus multiplicative
renormalizability  imposes the
following constraint~:
\begin{eqnarray}
 && \nonumber \quad  \frac{3}{2}\ \int_{0}^{p^2}
                \frac{dk^2}{p^2}\  \frac{k^2}{p^2}\
                \frac{k^2+p^2}{k^2-p^2}\
                \left( 1- \frac{F(k^2)}{F(p^2)}\ \right) \\ \nonumber
     && + \; \frac{3}{2}\ \int_{p^2}^{\Lambda^2}
           \frac{dk^2}{k^2}\
           \frac{k^2+p^2}{k^2-p^2}\
           \left( 1- \frac{F(k^2)}{F(p^2)}\ \right)   \\ \nonumber
     && +   \hspace{4 mm} \int_{0}^{p^2}\frac{dk^2}{p^2}\
             \frac{k^2}{p^2}\ F(k^2) \left( h_s(k^2,p^2) +
             h_a(k^2,p^2) \right)          \\
     && +  \hspace{4 mm} \int_{p^2}^{\Lambda^2}\frac{dk^2}{k^2}\
              F(k^2) \left( h_s(k^2,p^2) -
             h_a(k^2,p^2) \right) = 0
\end{eqnarray}
where $F(p^2)=A(p^{2})^{\nu}$ and the artificial cut-off, $\Lambda$,
 can be taken
to infinity with impunity. The scale invariance of the integrals makes
it convenient to introduce the variable $x$, where for $ 0 \le k^{2} <
p^{2} $, $x=k^{2}/p^{2}$, and for $ p^{2} \le k^{2} < \infty $,
$x=p^{2}/k^{2}$ \cite{Maris,ABGPR}. Then

\vfil\eject

\begin{eqnarray}
 &&  \nonumber  \frac{3}{2}\    \int_{0}^{1}  dx\, \frac{x+1}{x-1}\
                 r_1(x)  \\  \nonumber
 && +  \int_{0}^{1}  dx \,x^{\nu+1} F(p^2)\left( h_s(xp^2,p^2) +
          h_a(xp^2,p^2) \right)  \\
 && +  \int_{0}^{1}  dx \,x^{-\nu-1} F(p^2)\left(
        h_s\left(p^2/x ,p^2  \right)  -
          h_a\left(p^2/x,p^2 \right) \right) = 0
\end{eqnarray}
where
\begin{eqnarray*}
            r_1(x)&=&x(1-x^{\nu})-x^{-1}(1-x^{-\nu})  \\
            r_1 \left({1/x}\right)&=&-r_1(x)\qquad .
\end{eqnarray*}
Since this equation must hold true at all $p^{2}$, the integrands
cannot be functions of  $p^{2}$ but only of $x$. Thus
\begin{eqnarray*}
            F(p^2)\,h_s(xp^2,p^2)& \equiv &h_1(x),   \\
            F(p^2)\,h_a(xp^2,p^2)& \equiv &h_2(x)
\end{eqnarray*}
defines $h_{1},h_{2}$. Then Eq. (26) becomes
\begin{eqnarray}
 \nonumber   \frac{3}{2}\   \int_{0}^{1}  dx \,\frac{x+1}{x-1}\
           r_1(x)
           &+&  \int_{0}^{1}  dx\, x^{\nu+1} \left( h_1(x)+h_2(x) \right)    \\
  &+&  \int_{0}^{1}  dx\, x^{-\nu-1} \left( h_1 \left( {1/x}
       \right) - h_{2}\left( {1/x} \right) \right) = 0\quad .
\end{eqnarray}
  The original symmetry of the $\tau$'s under the exchange of $k^2$
and $p^2$ translates as follows in terms of the $x$-variable \cite{Dong}~:
\begin{eqnarray*}
    h_1\left({1/x} \right)&=&\;x^{\nu}  h_1(x)          \\
    h_2\left({1/x} \right)&=&-x^{\nu} h_2(x)
\end{eqnarray*}
   In the most compact way, Eq. (27) can be written as~:
\begin{eqnarray}
       \int_{0}^{1}  dx \;W_1(x)=0 \qquad ,
\end{eqnarray}
where
\begin{eqnarray}
    W_1(x)= \frac{3}{2}\ \frac{x+1}{x-1}\ r_1(x) + \left( x^{\nu+1}+x^{-1}
           \right)  \left(  h_1(x) +  h_2(x) \right)\quad .
\end{eqnarray}
Thus this function $W_1(x)$ fixes $\tau_{6}(k^2,p^2)$ and the
combination  $\overline{\tau}(k^2,p^2)$, so that
\begin{eqnarray}
   \overline{\tau}(k^2,p^2)\;= && \;\ \frac{1}{4}\ \frac{1}{k^2-p^2}\
        \frac{1}{s_{1}(k^2,p^2)}\  \Bigg[
         W_{1}\left( \frac{k^2}{p^2}\ \right) - W_{1}\left(\frac{p^2}{k^2}\
         \right) \Bigg]
        \\ \nonumber
        \vspace{11mm} \\ \nonumber
\tau_{6}(k^2,p^2)\;=&&-\,\frac{1}{2}\ \frac{k^2+p^2}{(k^2-p^2)^2}
    \left( \frac{1}{F(k^2)} - \frac{1}{F(p^2)} \right)      +
      \frac{1}{3}\, \frac{k^2+p^2}{k^2-p^2}\ \overline{\tau}(k^2,p^2) \\
  &&+ \,\frac{1}{6}\ \frac{1}{k^2-p^2}\  \frac{1}{s_{1}(k^2,p^2)}\
         \Bigg[ W_{1}\left(\frac{k^2}{p^2}\ \right)  +
        W_{1}\left(\frac{p^2}{k^2}\ \right)  \Bigg]
\end{eqnarray}
where
\begin{eqnarray*}
 s_{1}(k^2,p^2)= \frac{k^2}{p^2}\ F(k^2) +  \frac{p^2}{k^2}\ F(p^2)\quad.
\end{eqnarray*}
It is the first term in Eq. (31) that is essentially the {\em CP} vertex in the
massless theory. Note the automatic appearance of the difference
$\left({F(k^2)}^{-1}-{F(p^2)}^{-1}\right)$, which Curtis et al.
\cite{CP} conjectured was the non-perturbative generalization of the leading
logarithm behaviour in lowest order perturbation theory, Eq. (11).
Indeed, agreement with this behaviour is naturally achieved if
$W_1 \to 0$ in this limit.

 The vertex can only have singularities for good dynamical reasons.
It cannot have kinematic singularities. A sufficient condition
for this is to assume that
 each of the $\tau_i$ (i=1,8) is free of
kinematic singularities.   Ball and Chiu \cite{BC} found that
with their choice of basis vectors $T_i^{\mu}$ this is indeed true
at one loop order in perturbation theory in the Feynman gauge and
 Kizilers\" u \cite{Ayse}
has shown this in any covariant gauge at this order too.
In the present non-perturbative analysis
that this continues to hold with the Ball-Chiu basis vectors
is a plausible  simplifying
assumption. Thus
\begin{eqnarray}
 \lim_{k^{2} \rightarrow  p^{2} }
(k^{2}-p^{2})\, \tau_{6}(k^{2},p^{2})=0\quad ,
\end{eqnarray}
which requires
\begin{eqnarray}
     W_{1}(1)+W'_{1}(1)=-6\nu\qquad ,
\end{eqnarray}
as found by \cite{Dong}.
Perturbation theory demands $W_1(x)$ be ${\cal O}(\alpha)$.
While the form of the coefficient function $\tau_6$
 is determined by the constrained function $W_1(x)$,
  it is only the combination $\overline{\tau}$ of $\tau_2,\, \tau_3,\,\tau_8$
that is so specified.  By imposing the gauge independence of the critical
coupling for
mass generation, we will be able to separate these functions as we now show in
Sect. 3.

\vfil\eject

\baselineskip=9.5mm
\section{The Mass Function}

While for $\alpha < \alpha_{c}$, there is only one solution
${\cal M}(p^2)=0$, as $\alpha \rightarrow \alpha_{c}$ a second non-zero
solution becomes possible. This solution bifurcates away from the
other solution. Bifurcation analysis allows us to
investigate precisely when this happens. In the neighbourhood of the
critical coupling, terms quadratic in the mass function can be
rigorously neglected. Thus the wavefunction renormalization, $F(p^2)$,
is that of the massless theory, Sect. (2), and the equation for the
mass function, ${\cal M}(p^2)$, Eq. (13) with $m_0 \equiv 0$, linearizes~:
\vspace {3mm}
\begin{eqnarray}
 \nonumber \frac{{\cal M}(p^2)}{F(p^2)}\ &=& \frac{\alpha\xi}{4\pi}\
                          \int_{0}^{p^2}
                          \frac{dk^2}{p^2}\ {\cal M}(k^2)
                          \frac{F(k^2)}{F(p^2)}\ +
                          \frac{\alpha\xi}{4\pi}\
                          \int_{p^2}^{\Lambda^2}
                          \frac{dk^2}{k^2}\ {\cal M}(p^2)
                          \frac{F(k^2)}{F(p^2)}\     \\  \nonumber
                   &+& \frac{3\alpha}{4\pi}\
                         \vspace{5 mm} \int_{0}^{p^2}\frac{dk^2}{p^2}\
                         \Bigg[  {\cal M}(k^2) +
                         \frac{p^2}{2(k^2-p^2)}\ {\cal M}(k^2)
                         \left( 1- \frac{F(k^2)}{F(p^2)}\
                         \right)              \\  \nonumber
                   &&  \hspace{30 mm}\ -
                          \frac{k^2}{2(k^2-p^2)}\,
                                 \left({\cal M}(k^2) -
                         {\cal M}(p^2)\frac{F(k^2)}{F(p^2)}\
                         \right)  \Bigg]  \\ \nonumber
                   &+& \frac{3\alpha}{4\pi}\
                         \vspace{5 mm} \int_{p^2}^{\Lambda^2}
                         \frac{dk^2}{k^2}\
                         \Bigg[ {\cal M}(k^2) \frac{F(k^2)}{F(p^2)} +
                         \frac{k^2}{2(k^2-p^2)}\ {\cal M}(k^2)
                         \left( 1- \frac{F(k^2)}{F(p^2)}\
                         \right)              \\  \nonumber
                   &&  \hspace{30 mm}\ -
                          \frac{p^2}{2(k^2-p^2)}\,
                         \left({\cal M}(k^2) -
                         {\cal M}(p^2)\frac{F(k^2)}{F(p^2)}\
                         \right)  \Bigg]  \\ \nonumber
                   &-&  \frac{3\alpha}{4\pi}\
                         \vspace{5 mm} \int_{0}^{p^2}
                         \frac{dk^2}{p^2}
                         {\cal M}(k^2) F(k^2)
                         \Bigg[ \frac{k^2}{6} (k^2-3p^2)
                          \tau_{2}(k^2,p^2) \\
                         \nonumber
                   && \hspace{30 mm}\ +\,  p^2 \,\tau_{3}(k^2,p^2)
                        \, + \, (k^2-p^2)\,\tau_{6}(k^2,p^2)
                             \Bigg] \\ \nonumber
                   &-&  \frac{3\alpha}{4\pi}\
                         \vspace{5 mm} \int_{p^2}^{\Lambda^2}
                         \frac{dk^2}{k^2}
                          {\cal M}(k^2)F(k^2)
                           \Bigg[  \frac{p^2}{6} (p^2-3k^2)
                          \tau_{2}(k^2,p^2)    \\
                   &&  \hspace{30 mm}\ +\, k^2 \, \tau_{3}(k^2,p^2)
                         \,+ \, (k^2-p^2) \,\tau_{6}(k^2,p^2)
                             \Bigg] \, .
\end{eqnarray}
\vspace {3mm}

\noindent If this equation is to be multiplicatively renormalizable  with a
gauge independent bifurcation, then this imposes
further constraints on the transverse vertex, $\tau_i$ $(i=2,3,6)$.
We first work in the Landau gauge, where we assume the transverse
vertex vanishes. This is motivated by the perturbative result of Eq. (11).

 \noindent Then we have simply~:
 \vspace {2mm}
\begin{eqnarray}
 \nonumber {\cal M}(p^2) &=&  \frac{3\alpha}{4\pi}\
                         \vspace{5 mm} \int_{0}^{p^2}\frac{dk^2}{p^2}\
                         \Bigg [  {\cal M}(k^2) -
                          \frac{k^2}{2(k^2-p^2)}\
                         \left({\cal M}(k^2) -
                         {\cal M}(p^2)
                         \right)  \Bigg ] \\
                   &+&  \frac{3\alpha}{4\pi}\
                         \vspace{5 mm} \int_{p^2}^{\Lambda^2}
                         \frac{dk^2}{k^2}\
                         \Bigg [ {\cal M}(k^2) -
                          \frac{p^2}{2(k^2-p^2)}\
                         \left({\cal M}(k^2) -
                         {\cal M}(p^2)
                         \right)  \Bigg ] \, .
\end{eqnarray}

\vspace {2mm}
\noindent This equation has the multiplicatively renormalizable solution,
\begin{eqnarray}
      {\cal M}(k^2)=B \left( k^2 \right)^{-s}\qquad ,
\end{eqnarray}
where Eq. (35) requires,
\vspace {2mm}
\begin{eqnarray}
\frac{8 \pi}{3 \alpha}&=& 1 + \frac{3}{s} + \frac {1}{1-s} - \pi \cot
     \pi s  \;\equiv \; f(s)\; .
\end{eqnarray}

\noindent There are two roots for $s$ between 0 and 1.
Bifurcation occurs when the two roots for $s$ merge at $s=s_c$,
 specified by $f^{\prime}(s_c) = 0$.
  This point defines the critical coupling \cite{Atkinson,Maris,ABGPR},
  $\alpha_c = 8\pi/3 f(s_c)$.
 Numerically, $\alpha_{c} = 0.933667$ and $s_c = 0.470966$.
A little away from this critical point the exponent $s$ in Eq. (36)
is given by
\begin{eqnarray}
s\;=\;s_c\,
\pm\,\sqrt{{2f(s_c)\over{f^{\prime\prime}(s_c)}}}\,
\sqrt{1-{\alpha\over{\alpha_c}}}\; .
\end{eqnarray}

It is only at the bifurcation point that the simple behaviour of Eq.
(36) holds at all momenta. There, only when the mass is still
effectively zero is there just one scale, $\Lambda$,
for the momentum
dependence of ${\cal M}(k^2)$.  Multiplicative renormalizability then
forces a simple power behaviour. Such a multiplicatively
renormalizable mass function must exist in all gauges. Consequently,
 the exponent,
$s_c$, must be gauge independent.  Moreover, dynamical mass
 generation marks a physical phase change and so the critical coupling,
 $\alpha_c$, must also be gauge independent. Thus
the critical values, $\alpha_{c}$,  $s_{c}$, found in the
Landau gauge must hold in all gauges.
This is achieved as follows.
We recall Eqs. (14,25)~:
\begin{eqnarray}
 \frac{1}{F(p^2)}\ = 1+\frac{\alpha\xi}{4\pi}\
                          \int_{p^2}^{\Lambda^2}
                          \frac{dk^2}{k^2}\
                          \frac{F(k^2)}{F(p^2)} \qquad .
\end{eqnarray}

\vfil\eject

\noindent Multiplying this equation by ${\cal M}(p^2)$ and subtracting it from
Eq. (34), we obtain~:
\vspace {2mm}
\begin{eqnarray}
 \nonumber {\cal M}(p^2) &=& \frac{\alpha\xi}{4\pi}\
                          \int_{0}^{p^2}
                          \frac{dk^2}{p^2}\ {\cal M}(k^2)
                          \frac{F(k^2)}{F(p^2)}\  \\  \nonumber
                   &+& \frac{3\alpha}{4\pi}\
                         \vspace{5 mm} \int_{0}^{p^2}\frac{dk^2}{p^2}\
                         \Bigg[  {\cal M}(k^2) +
                         \frac{p^2}{2(k^2-p^2)}\ {\cal M}(k^2)
                         \left( 1- \frac{F(k^2)}{F(p^2)}\
                         \right)              \\  \nonumber
                   &&  \hspace{30 mm}\ -
                          \,\frac{k^2}{2(k^2-p^2)}\
                         \left({\cal M}(k^2) -
                         {\cal M}(p^2)\frac{F(k^2)}{F(p^2)}\
                         \right)  \Bigg]  \\ \nonumber
                   &+& \frac{3\alpha}{4\pi}\
                         \vspace{5 mm} \int_{p^2}^{\Lambda^2}
                         \frac{dk^2}{k^2}\
                         \Bigg[  {\cal M}(k^2) \frac{F(k^2)}{F(p^2)} +
                         \frac{k^2}{2(k^2-p^2)}\ {\cal M}(k^2)
                         \left( 1- \frac{F(k^2)}{F(p^2)}\
                         \right)              \\  \nonumber
                   &&  \hspace{30 mm}\ -
                         \, \frac{p^2}{2(k^2-p^2)}\
                         \left({\cal M}(k^2) -
                         {\cal M}(p^2)\frac{F(k^2)}{F(p^2)}\
                         \right)  \Bigg]  \\ \nonumber
                   &-&  \frac{3\alpha}{4\pi}\
                         \vspace{5 mm} \int_{0}^{p^2}
                         \frac{dk^2}{p^2}
                        {\cal M}(k^2)F(k^2)
                        \Bigg[ \frac{k^2}{6} (k^2-3p^2)
                        \tau_{2}(k^2,p^2) \\
                         \nonumber
                   &&  \hspace{30 mm}\ + \, p^2 \,\tau_{3}(k^2,p^2)
                        \, +\,(k^2-p^2) \, \tau_{6}(k^2,p^2)
                            \Bigg] \\ \nonumber
                   &-&  \frac{3\alpha}{4\pi}\
                         \vspace{5 mm} \int_{p^2}^{\Lambda^2}
                         \frac{dk^2}{k^2}
                         {\cal M}(k^2)F(k^2)
                           \Bigg[ \frac{p^2}{6} (p^2-3k^2)
                       \,  \tau_{2}(k^2,p^2)\\
                   && \hspace{30 mm}\ +\, k^2 \,\tau_{3}(k^2,p^2)
                       \,  +\, (k^2-p^2) \,\tau_{6}(k^2,p^2) \Bigg] \, .
\end{eqnarray}
\vspace {2mm}
\noindent In order for the above equation to reduce to Eq. (35), it must be
true that~:
\vspace {2mm}
\begin{eqnarray}
 \nonumber             \frac{\xi}{3} \int_{0}^{p^2}
                          \frac{dk^2}{p^2}\ {\cal M}(k^2)
                          \frac{F(k^2)}{F(p^2)}\ &=&
                     - \vspace{5 mm} \int_{0}^{p^2} dk^2 \
                          \frac{{\cal M}(k^2)}{2(k^2-p^2)}
                         \left( 1- \frac{F(k^2)}{F(p^2)}\
                         \right) \\  \nonumber
                   & &  - \vspace{5 mm} \int_{p^2}^{\Lambda^2}
                           dk^2 \
                          \frac{{\cal M}(k^2)}{2(k^2-p^2)}
                         \left( 1- \frac{F(k^2)}{F(p^2)}\
                         \right)  \\  \nonumber
                   & & + \vspace{5 mm} \int_{0}^{p^2}
                         \frac{dk^2}{p^2}
                         {\cal M}(k^2) F(k^2) \Bigg[ \frac{k^2}{6}
                         (k^2-3p^2)\,\tau_{2}(k^2,p^2) \\
                         \nonumber
                   & & \hspace{20mm} +\,  p^2 \,\tau_{3}(k^2,p^2)
                        \, +\,  (k^2-p^2)\,\tau_{6}(k^2,p^2)
                             \Bigg] \\ \nonumber
                   & & + \vspace{5 mm} \int_{p^2}^{\Lambda^2}
                         \frac{dk^2}{k^2}
                       {\cal M}(k^2) F(k^2) \Bigg[  \frac{p^2}{6}
                         (p^2-3k^2)\, \tau_{2}(k^2,p^2)  \\
                   & & \hspace{20mm} +\, k^2 \,\tau_{3}(k^2,p^2)
                         \,+ \, (k^2-p^2)\, \tau_{6}(k^2,p^2)  \Bigg]
\end{eqnarray}
\vspace {2mm}
\noindent at all momentum $p$ and in all gauges $\xi$.

\vfil\eject

\noindent This equation can be written as follows~:
\vspace {2mm}
\begin{eqnarray}
 \nonumber             \xi \int_{0}^{p^2}
                          \frac{dk^2}{p^2}\, {\cal M}(k^2)
                          \frac{F(k^2)}{F(p^2)}\ &=&
                     - \vspace{5 mm} \int_{0}^{p^2} dk^2
                         \, \frac{3{\cal M}(k^2)}{2(k^2-p^2)}
                         \left( 1- \frac{F(k^2)}{F(p^2)}\
                         \right) \\  \nonumber
                   & &  - \vspace{5 mm} \int_{p^2}^{\Lambda^2}
                          dk^2  \,
                          \frac{3{\cal M}(k^2)}{2(k^2-p^2)}
                         \left( 1- \frac{F(k^2)}{F(p^2)}\
                         \right) \\  \nonumber
                   & & + \vspace{5 mm} \int_{0}^{p^2}
                         \frac{dk^2}{p^2}\,
                          {\cal M}(k^2)F(k^2)
                          K_{3}(k^2,p^2) \\
                   & & + \vspace{5 mm} \int_{p^2}^{\Lambda^2}
                         \frac{dk^2}{k^2} \,
                          {\cal M}(k^2)F(k^2)
                          K_{4}(k^2,p^2)\quad,
\end{eqnarray}
\vspace {2mm}
\noindent where $K_{3}(k^2,p^2)$ and $K_{4}(k^2,p^2)$ can, like
$K_{1}(k^2,p^2)$
and $K_{2}(k^2,p^2)$, be expressed in terms of functions with
definite symmetry properties under the interchange of $k$ and $p$~:
 \begin{eqnarray}
\nonumber  g_s(k^2,p^2)\,&=&\,\frac{1}{4}\ \left[ (k^2-p^2)^2 -4k^2p^2
              \right] \tau_2(k^2,p^2) + \frac{3}{2}\ (k^2+p^2)
              \,\tau_3(k^2,p^2)    \\ \nonumber
             && \hspace{21mm} +3(k^2-p^2) \,\tau_6(k^2,p^2) \\
           g_a(k^2,p^2)&=& \frac{1}{4} (k^2-p^2) \left[ (k^2+p^2)
           \tau_2(k^2,p^2) - 6 \tau_3(k^2,p^2) \right]
\end{eqnarray}
so that
\begin{eqnarray*}
   K_{3}(k^2,p^2)&=&g_s(k^2,p^2)+g_a(k^2,p^2)  \\
   K_{4}(k^2,p^2)&=&g_s(k^2,p^2)-g_a(k^2,p^2)\qquad .
\end{eqnarray*}
Introducing the variable $x$ as before and knowing that
${\cal M}(k^2)\,\sim\, (k^2)^{-s_c}$  and $F(k^2)\,\sim\,(k^2)^{\nu}$,
Eq. (42) becomes,
\vspace {2mm}
\begin{eqnarray}
 &&  \nonumber  \xi   \int_{0}^{1}  dx \,x^{\nu-s_c} + \frac{3}{2}\
                \int_{0}^{1}  \frac{dx}{x-1}\  \left[ x^{-s_c} -
                x^{\nu-s_c} - x^{s_c-1} + x^{s_c-\nu-1}
                \right]
                  \\  \nonumber
 && -  \int_{0}^{1}  dx \,x^{\nu -s_c}
        F(p^{2}) \left[ g_s(xp^{2},p^{2})+g_a(xp^{2},p^{2}) \right] \\
 && -   \int_{0}^{1}  dx \,x^{s_c-\nu -1}  F(p^{2})
         \left[ g_s\left(p^2/x,p^{2} \right) -
         g_a\left(p^2/x, p^{2} \right] \right) \qquad =\; 0 \quad .
\end{eqnarray}
\vspace {2mm}
\noindent Once again, this equation must hold true for all $p^{2}$, and so the
integrands cannot be functions of $p^{2}$ but solely of $x$. Thus, we
conveniently define,
\begin{eqnarray*}
            F(p^2)\,g_s(xp^2,p^2)& \equiv &g_1(x)   \\
            F(p^2)\,g_a(xp^2,p^2)& \equiv &g_2(x)\qquad .
\end{eqnarray*}
Then we have
\vspace {2mm}
\begin{eqnarray}
 &&  \nonumber  \xi   \int_{0}^{1}  dx\, x^{\nu-s_c} + \frac{3}{2}\
                \int_{0}^{1}  \frac{dx}{x-1}\  \left[ x^{-s_c} -
                x^{\nu-s_c} - x^{s_c-1} + x^{s_c-\nu-1}
                \right]
                  \\
 && -  \int_{0}^{1}  dx\, x^{\nu -s_c}
         \left[ g_1(x)+g_2(x) \right]
 -   \int_{0}^{1}  dx\, x^{s_c-\nu -1}
         \left[ g_1\left({1/x} \right)-g_2 \left({1/x}
         \right) \right] \;=\; 0\quad .
\end{eqnarray}
\vspace {2mm}
\noindent The symmetry of the vertex \cite{Dong} under $k \leftrightarrow p$
means that,
\begin{eqnarray*}
    g_1\left({1/x} \right)&=&\;x^{\nu}  g_1(x)          \\
    g_2\left({1/x} \right)&=&-x^{\nu} g_2(x)\qquad .
\end{eqnarray*}
In contrast to our discussion in Sect. 2 when the equations for the
wavefunction renormalization, $F(p^2)$, apply for all values of the coupling,
Eqs. (44,45) only hold when $\alpha = \alpha_c$.

 Eq. (45) can be written in a compact way as
\begin{eqnarray}
       \int_{0}^{1}  {dx\over{\sqrt{x}}} \; W_2(x)\;=\;0\qquad ,
\end{eqnarray}
where
\begin{eqnarray}
\nonumber    W_2(x)&=& \xi\ x^{\nu -s_c + \frac{1}{2}\ } + \frac{3}{2}\
            \frac{r_2(x)}{x-1} - x^{\nu -s_c + \frac{1}{2}\ }
           \left[ g_1(x)+g_2(x) \right]  \\
            &-& x^{-\nu +s_c - \frac{1}{2}\ }
           \left[ g_1\left({1/x} \right)-g_2\left({1/x}
           \right) \right]
\end{eqnarray}
with
\begin{eqnarray}
   r_2(x)= x^{ \frac{1}{2} -s_c}\ (1-x^{\nu}) - x^{s_c - \frac{1}{2}\ }\
           (1-x^{-\nu})
\end{eqnarray}
which has the property,  $r_2\left(1/x \right) = -r_2(x)$.
Conveniently defining the combination,
\begin{eqnarray}
     s_2(k^2,p^2)=\frac{k}{p}\,  \frac{{\cal M}(k^2)}{{\cal M}(p^2)}\,
             F(k^2) + \frac{p}{k}\,  \frac{{\cal M}(p^2)}{{\cal
             M}(k^2)}\, F(p^2)
\end{eqnarray}
we have
\begin{eqnarray}
\nonumber
 g_s(k^2,p^2)&=&\frac{\xi}{2s_2(k^2,p^2)}
          \Bigg[ \frac{k}{p}\,  \frac{{\cal M}(k^2)F(k^2)}{{\cal
           M}(p^2)F(p^2)}\,
            + \frac{p}{k}\,  \frac{{\cal M}(p^2)F(p^2)}{{\cal
             M}(k^2)F(k^2)}\, \Bigg] \\  \nonumber
 & &+\frac{3}{4}\,  \frac{k^2+p^2}{k^2-p^2} \frac{1}{s_2(k^2,p^2)}\; r_2\left(
            \frac{k^2}{p^2}\ \right) \\
 & &-\frac{1}{2}\
            \frac{1}{s_2(k^2,p^2)} \ \Bigg[ W_2\left(
            \frac{k^2}{p^2}\ \right) +  W_2\left( \frac{p^2}{k^2}\
            \right) \Bigg]       \, ,
\end{eqnarray}
\begin{eqnarray}
 \nonumber
 g_a(k^2,p^2)&=&\frac{\xi}{2s_2(k^2,p^2)}
          \Bigg[ \frac{k}{p}\,  \frac{{\cal M}(k^2)F(k^2)}{{\cal
           M}(p^2)F(p^2)}\,
            - \frac{p}{k}\,  \frac{{\cal M}(p^2)F(p^2)}{{\cal
             M}(k^2)F(k^2)}\,    \Bigg] \\ \nonumber
 & &- \frac{3}{4} \; \frac{1}{s_2(k^2,p^2)}\ \; r_2\left( \frac{k^2}{p^2}\
         \right) \\
 & &- \frac{1}{2}\ \frac{1}{s_2(k^2,p^2)}\
            \Bigg[ W_2\left( \frac{k^2}{p^2}\ \right) -  W_2\left(
          \frac{p^2}{k^2}\ \right)
            \Bigg] \, .
\end{eqnarray}
Solving the last two equations for $\tau_{2}$ and $\tau_{3}$ in terms
of $\tau_6$ and $W_2$, we obtain~:
\vspace {2mm}
\begin{eqnarray}
\nonumber \tau_2(k^2,p^2)&=& \frac{2\xi}{(k^2-p^2)^2} \,\frac{
          q_2(k^2,p^2)}{ s_2(k^2,p^2)} -6 \,\frac{ \tau_6(k^2,p^2)}{
          (k^2-p^2)} \\  \nonumber
       & &-\frac{1}{(k^2-p^2)^2}\  \frac{1}{s_2(k^2,p^2)}\
           \Bigg[ W_2\left( \frac{k^2}{p^2}\ \right) +  W_2\left(
            \frac{p^2}{k^2}\ \right) \Bigg]  \\
       & &-\frac{k^2+p^2}{(k^2-p^2)^3}\  \frac{1}{s_2(k^2,p^2)}\
           \Bigg[ W_2\left( \frac{k^2}{p^2}\ \right) -  W_2\left(
            \frac{p^2}{k^2}\ \right) \Bigg] \, ,
\end{eqnarray}
\vspace{2mm}
\noindent where
\begin{eqnarray}
  q_2(k^2,p^2)= \frac{1}{k^2-p^2} \Bigg[ \frac{k^3}{p}\,
              \frac{{\cal M}(k^2)F(k^2)}{{\cal M}(p^2)F(p^2)}
               -  \frac{p^3}{k}\,
              \frac{{\cal M}(p^2)F(p^2)}{{\cal M}(k^2)F(k^2)} \Bigg] \, ,
\end{eqnarray}
where $q_2(k^2,p^2)$ is  obviously a symmetric function of $k$ and $p$,
and
\vfil\eject
\vspace {2mm}
\begin{eqnarray}
\nonumber \tau_3(k^2,p^2)&=&  - \frac{k^2+p^2}{
          k^2-p^2}\, \tau_6(k^2,p^2) \\ \nonumber
          & &+ \frac{1}{k^2-p^2}\
          \frac{1}{s_2(k^2,p^2)}\ \Bigg[ \frac{1}{2}\ r_2\left(
          \frac{k^2}{p^2}\ \right) - \frac{\xi}{3}
          \,q_3(k^2,p^2) \Bigg]    \\ \nonumber
          & &- \frac{1}{6} \frac{k^2+p^2}{(k^2-p^2)^2}\
          \frac{1}{s_2(k^2,p^2)}\  \Bigg[ W_2\left( \frac{k^2}{p^2}\ \right) +
          W_2\left( \frac{p^2}{k^2}\ \right) \Bigg]  \\
          & & +\frac{1}{6} \frac{k^4+p^4-6k^2p^2}{(k^2-p^2)^3}\
          \frac{1}{s_2(k^2,p^2)}\  \Bigg[ W_2\left( \frac{k^2}{p^2}\ \right) -
          W_2\left( \frac{p^2}{k^2}\ \right) \Bigg] \, ,
\end{eqnarray}
\vspace {2mm}
\noindent where
\begin{eqnarray}
 q_3(k^2,p^2)=  \frac{kp}{(k^2-p^2)^2} \Bigg[ (p^2-3k^2)\,
               \frac{{\cal M}(k^2)F(k^2)}{{\cal M}(p^2)F(p^2)} - (k^2-3p^2)\,
               \frac{{\cal M}(k^2)F(k^2)}{{\cal M}(p^2)F(p^2)} \Bigg]  ,
\end{eqnarray}
where $q_3(k^2, p^2)$ is antisymmetric in $k$ and $p$.
The relation, Eq. (17),
\begin{eqnarray*}
    \overline{\tau}(k^2,p^2)&=&\tau_{3}(k^2,p^2) + \tau_{8}(k^2,p^2) -
                  \frac{1}{2}\ (k^2+p^2) \tau_{2}(k^2,p^2)
\end{eqnarray*}
then fixes $\tau_{8}(k^2,p^2)$.
\begin{eqnarray}
          \tau_8(k^2,p^2)&=&  -2\, \frac{k^2+p^2}{
          k^2-p^2} \,\tau_6(k^2,p^2) + \overline{\tau}(k^2,p^2) \\ \nonumber
          & &-  \frac{1}{k^2-p^2}\
          \frac{1}{s_2(k^2,p^2)}\ \Bigg[ \frac{1}{2}\ r_2\left(
          \frac{k^2}{p^2}\ \right) - \frac{\xi}{3}
           \,q_8(k^2,p^2)
          \Bigg]        \\  \nonumber
          & &- \frac{1}{3}\ \frac{k^2+p^2}{(k^2-p^2)^2}\
            \frac{1}{s_2(k^2,p^2)}\
           \Bigg[ W_2\left( \frac{k^2}{p^2}\ \right) +  W_2\left(
            \frac{p^2}{k^2}\ \right) \Bigg]    \\
          & &- \frac{2}{3}\ \frac{k^4+p^4}{(k^2-p^2)^3}\
            \frac{1}{s_2(k^2,p^2)}\
           \Bigg[ W_2\left( \frac{k^2}{p^2}\ \right) -  W_2\left(
            \frac{p^2}{k^2}\ \right) \Bigg]  \, ,
\end{eqnarray}
\vspace {2mm}
\noindent where
\begin{eqnarray}
 q_8(k^2,p^2)=  \frac{1}{(k^2-p^2)^2} \Bigg[\frac{k}{p}(3k^4+p^4)\,
                \frac{{\cal M}(k^2)F(k^2)}{{\cal M}(p^2)F(p^2)} -
                \frac{p}{k}(k^4+3p^4)\, \frac{{\cal
                M}(k^2)F(k^2)}{{\cal M}(p^2)F(p^2)}
                \Bigg] \ ,
\end{eqnarray}
which is clearly antisymmetric in $k$ and $p$.
Imposing the condition that the vertex and its components should be
free of kinematic singularities means that,
\begin{eqnarray*}
  \lim_{k^{2} \rightarrow p^{2}
  }(k^{2}-p^{2})\,\tau_{i}(k^{2},p^{2})=0
   \hspace{10 mm} \mbox{$i=2,3,8$} \quad ,
\end{eqnarray*}
noting that the antisymmetry of $\tau_6$ means $\tau_6(p^2,p^2)\,=\,0$.  Thus
\begin{eqnarray}
             W_{2}(1)+2W'_{2}(1)= 2\xi (\nu-s+1) \qquad ,
\end{eqnarray}
where $s=s_c$ at the critical point.
The transverse vertex has the correct lowest order perturbative limit,
viz. $\Gamma_T^{\mu} = {\cal O}(\alpha)$, provided,
\begin{eqnarray}
 W_{2}\left(k^2/p^2\right)\,=\,\xi\ \frac{k}{p}\ \frac{{\cal
         M}(k^2)F(k^2)}{{\cal M}(p^2)F(p^2)}\, + \,{\cal O}(\alpha) \quad.
\end{eqnarray}
Since at large momenta we expect the power behaviour of Eqs. (24,36)
even away from criticality, Eq. (59) will hold for all values of the
coupling, $\alpha$. In contrast, Eq. (46) is only true at the
bifurcation point. Its exact form for all $\alpha$
is not known, but Eq. (38) might suggest
\begin{eqnarray}
       \int_{0}^{1}  {dx\over{\sqrt{x}}} \; W_2(x)\,\approx\,\xi\,
       \sqrt{1-{\alpha\over{\alpha_c}}}\quad ,
\end{eqnarray}
to agree with  both the $\alpha = 0$ and
 $\alpha = \alpha_c$ limits, Eqs. (60,46).
These equations determine our vertex for any $W_i$
$(i=1,2)$ that satisfy the constraints.

 In \cite{ABGPR} a plot is shown of the
critical coupling, $\alpha_c$, as a function of the covariant gauge parameter,
$\xi$, when the {\em CP} vertex is used. We refrain from showing the analogous
graph for the presently
constructed vertex, as $\alpha_c$ would be boringly gauge independent !
This has been achieved for any choice of the functions $W_i(x)$
$(i=1, 2)$, that satisfy Eqs. (28, 33, 46, 59, 60).
A simple example for $W_1$ is
$2\nu\ (1-2x)$. There are, of course, an infinity of such functions.
 In practice,
we expect that $W_1$ should be expressible solely in terms of the ratio
$F(k^2)/ F(p^2)$, while $W_2$ should surely also involve
${\cal M}(k^2)/{\cal M}(p^2)$. However, we have
not been able to find  simple examples that achieve this.
 The exact form of the full vertex would, of course, determine these functions
 precisely.
 Thus solving the Schwinger-Dyson equation   ~for the three point function
would specify the unknowns.
 However, that has not been our aim.  Our aim is more limited.
  It is to construct a vertex that ensures the fermion propagator is
  gauge covariant, multiplicatively renormalizable and has a gauge independent
chiral symmetry breaking
  phase transition.
One does not need to know the exact form of the full vertex to achieve these
properties only the {\em effective} vertex for the fermion equation, Eq. (1).
However, we believe that  this effective vertex should nevertheless
satisfy the appropriate Ward-Takahashi identity and agree with
 perturbation theory at least in the leading logarithmic limit of
the weak coupling regime.
 This is the construction we have achieved for any functions
 $W_i(x)$ $(i=1, 2)$. This {\em effective} vertex is thus given by Eqs.
 (5,6,7,9,10,31,51,53,56).

\vskip 1cm
\section{Conclusions}
\baselineskip=9.5mm

The non-perturbative behaviour of
the fermion propagator is governed by its Schwinger-Dyson equation  .  In
quenched QED,
the self-consistent solution of this equation is determined by the
fermion-boson interaction.
This in turn satisfies a Schwinger-Dyson equation  ~that relates
it to the full 4-point function and this
4-point function satisfies its own Schwinger-Dyson equation
 While the solution of this infinite set of
 equations represents the whole theory,
the complete set is, of course, impossible to solve.  Consequently, we need a
systematic method of truncation that maintains the key features of the theory :
its gauge invariance and multiplicative renormalizability.
The only known truncation scheme consistently respecting these
properties is perturbation theory.  However, the bulk of strong interaction
phenomena require a non-perturbative approach.
Thus, for example, massless bare matter
fields remain massless to all orders in perturbation theory. However, if the
interactions are strong enough, a chiral symmetry breaking phase may become
 a possibility.
  Truncating the nested Schwinger-Dyson equations to just the
 fermion equation by the {\it rainbow} approximation, in which
 the fermion-boson vertex is simply treated as bare,
 this possibility is realized.
 However, this approximation is highly gauge dependent with
 the critical coupling for this phase transition varying by a factor of two
 from $\xi=0$ to 3 \cite{CP3}.  The
 present paper defines a truncation of the fermion Schwinger-Dyson equation ,
 which does respect the key properties of the theory.
 The vertex constructed satisfies the Ward-Takahashi identity,
 ensures the fermion
 propagator is multiplicatively renormalizable,
  agrees with one loop perturbation
 theory for large momenta and enforces a gauge independent chiral symmetry
 breaking phase transition.
 This is a step on the way to a meaningful non-perturbative truncation scheme :
 meaningful in the sense that the fundamental aspects of the physics
 crucially determining the fermion propagator are thereby
 encoded in its Schwinger-Dyson equation.

  Investigation of how, for a given coupling strength, the generated mass
 compares with that found using the {\it rainbow} approximation requires
  the solution of the coupled equations for
  $F(p^2)$ and ${\cal M}(p^2)$.  Study of the chiral symmetry breaking phase
  transition, using bifurcation analysis,
  fortunately allows these equations to be uncoupled rigorously.
    The coupled solution is planned.

   The fact that in a (more) realistic version of non-perturbative QED,
   mass generation is possible makes it more,
   rather than less, likely that such a
   phase transition has been observed in heavy ion collisions \cite{Caldi}.
Moreover, it motivates
   the need for a realistic calculation of $t{\overline t}$ condensates
   as the source of the electroweak symmetry breaking \cite{Nambu}. A
realistic calculation, of course, requires the unquenching of the
theory. This brings at once renormalizations of the transverse photon
propagator and of the fermion-boson coupling. It is this renormalized
coupling, which at the corresponding chiral symmetry breaking phase
transition, is the physical quantity that must be gauge-independent.
The need to ensure the multiplicative renormalizability of the now
coupled photon propagator, of the fermion-boson coupling, as well as of
the fermion propagator, significantly complicates the problem.
The fermion-boson vertex (in particular its transverse part) will
intimately depend on the photon renormalization function in a
non-perturbative way not yet understood.

 Thus the complete multiplicative renormalizability of 2 and 3-point
functions brings not merely greater algebraic but also methodological
complexity. The results for quenched QED presented here provide the
starting point for such an investigation of full QED. The solution to
this problem will in turn be the starting point for a study of QCD,
where boson self-interactions, so essential for both asymptotic freedom
and confinement, will further complicate the analysis whether in
covariant or axial gauges. All this is for the future.

   \vskip 1cm

\noindent{\bf Acknowledgements}

   \vskip 5mm
   We are most grateful to our colleagues Jacques Bloch and Ay\c se Kizilers\"
u
   for lengthy  discussions at the start of this work
   and David Atkinson for his  useful comments
   on the draft version of this paper.
   AB wishes to thank the Government of Pakistan for a research studentship.

\vfil\eject
\hsize=16.5cm
\baselineskip=6mm

\vfil\eject
\vskip 1cm
\noindent{\bf Figure Caption}
\vskip 5mm
\noindent{ Fig. 1.} Schwinger-Dyson equation for the fermion propagator.

\hspace{8mm} The straight lines represent fermions and the wavy line
the photon.

\hspace{8mm} The solid dots
indicate full, as opposed to bare, quantities.

\end{document}